\documentstyle[12pt,aps,prb]{revtex}

\input psfig.tex

\begin{document}
\draft
\tightenlines

\title{Free energy and molecular dynamics calculations for the
cubic-tetragonal phase transition in zirconia}
\author{Stefano Fabris,\footnote{Present address: Max Planck Institut
F\"ur Metallforschung, Seestrasse 92, D-70174 Stuttgart, Germany.} Anthony
T. Paxton, and Michael W. Finnis} 
\address{Atomistic Simulation Group, Department of Pure and
Applied Physics, Queen's University, \\ Belfast BT7 1NN, United Kingdom}
\date{\today} 
\maketitle
\begin{abstract}
 The high-temperature cubic-tetragonal phase transition of pure
 stoichiometric zirconia is studied by molecular dynamics (MD) simulations
 and within the framework of the Landau theory of phase
 transformations. The interatomic forces are calculated using an empirical,
 self-consistent, orthogonal tight-binding (SC-TB) model, which includes
 atomic polarizabilities up to the quadrupolar level. A first set of
 standard MD calculations shows that, on increasing temperature, one
 particular vibrational frequency softens. The temperature evolution of the
 free energy surfaces around the phase transition is then studied with a
 second set of calculations. These combine the thermodynamic integration
 technique with constrained MD simulations. The results seem to support
 the thesis of a second-order phase transition but with unusual, very
 anharmonic behaviour above the transition temperature.
\end{abstract}


\section{Introduction}

 A large class of advanced ceramics are solid solutions of zirconia
 (ZrO$_2$) with cubic stabilising oxides like Y$_2$O$_3$, MgO or CeO, and
 are generally called {\it stabilized zirconias}. The long list of
 functional applications includes high-temperature devices, thermal
 barriers, and oxygen sensors. Moreover, partially stabilized zirconias
 represent a new generation of structural materials, by far the toughest
 ceramic oxides, strengthened by the mechanism called {\it transformation
 toughening}. The processing and service conditions of these materials
 involve phase transformations whose underlying physics is still
 a subject of controversy. One of these is the high-temperature
 cubic-tetragonal  phase transition which is the subject of the present
 paper. 

\begin{figure}
\caption{Cubic and tetragonal structures of ZrO$_2$. Light and dark
circles denote oxygen and zirconium atoms respectively. Arrows represent
the structural instability of the oxygen sublattice along the $X_2^-$
mode of vibration.}
\label{cell}
\centerline{\psfig{file=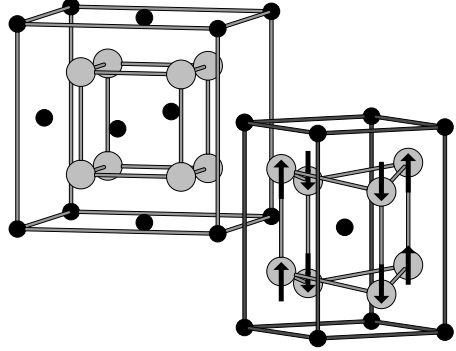,width=6cm,angle=-90}} 
\end{figure}

 Zirconia is monoclinic ($m$) at low
 temperatures,~\cite{Adam59,McCullough59,Smith65} tetragonal ($t$) between
 1400 and 2570 K,~\cite{Howard88,Teufer62} and cubic ($c$) up to the
 melting point of 2980 K.~\cite{Aldebert85,Ackermann77} High-temperature
 X-ray experiments on stabilized zirconia revealed the existence of a $c
 \leftrightarrow t$ phase transition between 2300 and 2600
 K,~\cite{Smith62,Wolten63b,Wolten63,Ruh67} depending on the atmosphere,
 but the mechanism of the transformation still has not been fully
 explained. The $c$ and $t$ unit cells are shown in Figure~\ref{cell}:
 note the characteristic tetragonal distortion of the oxygen sublattice
 in the $t$ phase.

 It is not possible to quench to low temperature the $c$ and $t$ forms of
 pure zirconia, hence the experiments are difficult because of the high
 temperatures involved. Alternatively, the $c$ and $t$ structures may be
 stabilized at low temperatures by impurities. The available measurements
 are mostly done on stabilized samples. This simplifies the experimental
 procedure but complicates the interpretation of the results, because,
 besides the equilibrium $t$ phase, other metastable tetragonal
 structures are observed in stabilized crystals, denoted by $t^\prime$
 and $t^{\prime \prime}$. The former is the microstructure of a solid
 solution quenched from the field of stability of the $c$ phase into the
 biphasic $c+t$ one.~\cite{Miller81,Yoshimura90} The $t$ and $t^\prime$
 forms are the same phase, they belong to the space group $P4_2/nmc$,
 but have different composition;~\cite{Yashima93} $t^\prime$ is also
 called {\it non transformable} because it does not spontaneously
 transform to the $m$ phase. The $t^{\prime \prime}$ structure is
 observed in the ZrO$_2$-ErO$_{1.5}$~\cite{Yashima93} and
 ZrO$_2$-Y$_2$O$_3$~\cite{Zhou91} systems, and has a cubic unit cell with
 the oxygen sublattice tetragonally distorted.

 The microstructure of samples rapidly cooled from the $c$-phase region
 presents twinned domains separated by antiphase boundaries. The nature
 and composition of these domains is related to the phase transition
 mechanism and has been a subject of controversy. Originally they have
 been interpreted as the result of a diffusionless martensitic
 reaction.~\cite{Cohen63,Scott75,Sakuma85,Michel83} Later, Heuer and
 R\"{u}hle~\cite{Heuer84} suggested that the transformation could be
 $non$-martensitic: homogeneous, massive, and displacive. Similarly, the
 observations of Lantieri {\it et al.}~\cite{Lantieri86} were interpreted
 to mean that the $c \rightarrow t^\prime$ transformation is
 diffusionless but non martensitic, and that the transformation always
 goes to completion. The same authors later proposed that the transition
 could be heterogeneous of the first-order with
 nucleation.~\cite{Heuer87} According to Sakuma,~\cite{Sakuma87} the
 transition is instead second-order.

 The temperature evolution of the tetragonality $c/a$ and of the anion
 sublattice distortion has been followed by Yashima {\it et
 al.}~\cite{Yashima93} in the ZrO$_2$-ErO$_{1.5}$ system.
 They showed that both order parameters depend continuously on the
 temperature and suggested that ``the transition has the nature of a
 higher-order phase transition''.

 Several attempts have been made in order to include this transformation in
 phenomenological theories. Hillert and Sakuma~\cite{Hillert91} expanded
 the free energy in terms of the defect concentration and assumed the
 transition to be second-order. Fan and Chen~\cite{Fan95} used the
 time-dependent Ginzburg-Landau theory to expand the free energy of the
 transformation, treating it as a first-order one. The transformation was
 instead assumed to be second-order in the Landau energy expansion of
 Katamura and Sakuma.~\cite{Katamura97}

 The theoretical treatment of the $c \leftrightarrow t$ transition is
 simpler in stoichiometric zirconia: this is a case similar to the
 $c\leftrightarrow t$ phase transition in BaTiO$_3$, where, according to
 symmetry considerations, the transformation could be either first or
 second order. A free energy Landau expansion for zirconia, involving the
 tetragonality of the cell only, without the distortion of the oxygen
 sublattice, inevitably predicts a first-order
 transformation.~\cite{Chan88} But the inclusion of the latter in the
 Landau expansion opens the possibility for a second-order
 transition.~\cite{Ishibashi89} As already pointed out~\cite{strcZrO2}
 the coupling between the order parameters may change the order of the
 transition from second to first.

 In the case of BaTiO$_3$ it has been possible to measure the order
 parameters very close to the transition
 temperature~\cite{Fleury71,Peercy72} and to establish the order and the
 mechanism of the transformation. Analogous experiments are difficult in
 pure zirconia because of the high transformation temperature.  The
 neutron diffraction analysis of Aldebert and Traverse~\cite{Aldebert85}
 provides the most complete thermomechanical description of pure $c$ and
 $t$ zirconia at high temperature. Aldebert and Traverse observe that:
 (i) The tetragonal distortion of the oxygen sublattice persists in the
 whole field of stability of the $t$ structure. (ii) The tetragonal
 distortion of the oxygen ions vanishes in the $c$ structure. (iii) The
 volume thermal expansion is linear and very close to isotropic up to
 near the transition point. As a consequence the $c/a$ ratio is almost
 temperature-independent over a wide range of temperatures, and sharply
 decreases near the transition temperature. (iv) The isotropic
 Debye-Waller factors of both species strongly increase before the
 transition temperature: the authors interpret it as a possible
 structural phenomenon anticipating the phase transition which could
 increase the ionic mobility.

 The plan of the present paper is as follows. In Sec.~\ref{theory} we
 introduce the main theoretical tools we have used to study the phase
 transition: the Landau theory of phase transformations, the thermodynamic
 integration technique, the constrained dynamics and the analysis of the
 order parameter fluctuations. The results are discussed in
 Sec.~\ref{SecRslt}. The phase transition mechanism was investigated using
 two sets of calculations. The first one, described in Sec.~\ref{SecMDrslt},
 is a traditional Molecular Dynamics (MD) analysis with which we observed the
 softening of a particular vibrational frequency. The second one, described
 in Sec.~\ref{SecFreeSurf}, combines the thermodynamic integration
 technique with constrained MD simulation to calculate the free energy
 surfaces around the phase transformation. We summarize the results in the
 final section.

\section{Theory} \label{theory}

\subsection{Landau theory of the phase transition} \label{SecLand}
 
 \subsubsection{Order parameters}
 
 The Landau theory of phase transformations~\cite{Landau5} describes the
 relationship between two crystal structures which share a common
 symmetry group {\boldmath $G$}$_{\rm 0}$. The disappearance of a
 particular symmetry operation is quantitatively described by order
 parameters, which are zero in the high symmetry phase and become
 non-zero in the low symmetry one. 

 Our preliminary analysis~\cite{strcZrO2} of the $c \leftrightarrow t$
 phase transformation, based on 0 K calculations, showed that the
 transformation is driven by the distortion of the anion sublattice which
 is described by the {\it primary} order parameter $\delta$. This is a
 measure of the distance between each oxygen atom and the corresponding
 centrosymmetric position it occupied in the $c$ structure. The $T=0$ K
 calculations of certain phonon frequencies of the $c$ phase show that a
 frequency of vibration at the $X$ point of the BZ is
 imaginary.~\cite{strcZrO2,Parlinski97,Detraux98} This phonon, labelled
 $X_2^-$, involves the oxygen sublattice only, and is shown in
 Figure~\ref{cell}. It transforms according to the $A_2$ irreducible
 representation of the little co-group of the $X$-point $D_{4h}$. The star
 of $D_{4h}$ contains three equivalent points, consequently the order
 parameter describing the tetragonal distortion has three components:
 $\delta_x$, $\delta_y$, and $\delta_z$.

 In transforming to the $t$ phase, the primitive unit cell doubles, so that
 the phonon corresponding to $X_2^-$ is at the $\Gamma$ point, and is
 generally labelled $A_{1g}$. Nevertheless, in order to unify the
 description for both the $c$ and the $t$ structures, we will not use this
 convention and we will always refer to the soft mode as the $X_2^-$ one,
 also in the $t$ phase.

 Besides the tetragonal distortion of the oxygen sublattice, it is
 necessary to capture the change of the unit cell shape. This is done by
 introducing auxiliary order parameters $\eta_i$ defined in terms of the
 strain tensor {\boldmath $ \epsilon $}. We decompose the six independent
 components of the strain tensor into irreducible representation of the
 $O_h$ cubic point group in Table~\ref{op-irrep}.

 It was shown previously that, at equilibrium, each auxiliary order
 parameter is second-order in $\delta$.~\cite{strcZrO2,Finnis} From now on,
 the order of the expansion terms will be expressed with respect to the
 order in $\delta$, therefore, as an example, a term like $\delta_x^2
 \eta_1$ is fourth order.

\subsubsection{Energy expansion}

 The Landau theory assumes that the appropriate thermodynamic potential of
 the crystal $\Phi$ can be expanded in powers of the order parameters about
 the transition point.  The Taylor expansion of $\Phi$ must be invariant
 under the symmetry operations of the high symmetry phase. As a
 consequence, the allowed terms in the expansion have to be
 symmetry-invariants as well, and can be found using group theory. The
 terms in the energy expansion will be polynomials in the strains $\epsilon_i$
 and displacements $\delta_i$ of Table~\ref{op-irrep}. We constructed all
 the possible polynomials up to the sixth-order and symmetrized them with
 respect to the symmetry operations of the cubic point group $O_h$. The
 resulting invariants are shown in Table~\ref{polnm}.

 This analysis showed that all the third-order invariants are identically
 zero, which is a necessary (but not sufficient) condition for a phase
 transition to be second-order. We already mentioned that the instability
 appears at the boundary of the BZ, therefore it halves the number of
 symmetry elements, and this is a further condition allowing the $c
 \leftrightarrow t$ phase transition to be second-order.~\cite{Landau5}

 In order to keep the discussion as simple as possible, at this stage we
 {\it assume} the phase transition to be second-order, truncating the
 Taylor expansion at the fourth-order term in $\delta$. The possible
 importance of the higher order terms will be discussed later. The
 energy expansion, expressed in terms of the basis function defined in
 Table~\ref{polnm}, is as follows:

\begin{eqnarray}
F & = & F_0 + \frac{a_2}{2} A_2 \! \left( \delta^2 \right) 
  + \sum_{i=1}^2  \frac{a_{4i}}{4} A_{4i} \! \left(\delta^4 \right)
  \label{energy} \\
  &   &  + \sum_{i=1}^3  b_{i} B_{i} \! \left(\epsilon,\delta^2 \right) 
  + \sum_{i=1}^3 \frac{c_{i}}{2}  C_{i} \! \left(\epsilon^2 \right) 
  + {\cal O}\left (\delta^6 \right). \nonumber 
\end{eqnarray}
 $F_0$ is the energy of the high-symmetry phase and is a function of the
 hydrostatic strain $\eta_1=Tr(${\boldmath $\epsilon$}$)$. The choice of
 the reference volume fixes $F_0$ and the expansion coefficients $a_2, \,
 \dots, c_3$. In the present case, the energy $F$ was expanded about the
 minimum of the energy-volume curve for the $c$ structure predicted by the
 SC-TB calculations.~\cite{strcZrO2}

 \subsection{Free energy calculation} \label{SecConstrMD}

 Free energy surfaces may be calculated directly from MD simulations in
 terms of {\it ensemble} averages by using the thermodynamic integration
 technique.~\cite{Frenkel96,Allen87} Here we briefly describe how this
 method was applied to zirconia.

 The thermodynamic integration method allows us to calculate free energy
 differences between a reference state, for which the internal energy $U_0$
 is known, and another state of the same system, with internal energy
 $U$. The idea is to relate the two structures with a {\it switching}
 parameter $\gamma$ which is zero in the reference state and non-zero
 otherwise. The free energy variation in the infinitesimal change
 $d\gamma$ may be calculated using standard statistical mechanics:
\begin{equation}
dF=\left\langle \frac{\partial U(\gamma)}{\partial \gamma}
\right\rangle_{\bar{\gamma}} d\gamma \label{therm-int}
\end{equation} 
 This is equivalent to the reversible work done for the structural
 modification described by $d\gamma$, implicitly assumed to be
 adiabatic. By $\left\langle \cdots \right\rangle$ we indicate the ensemble
 average, which has to be calculated at a constant value of
 $\gamma=\bar{\gamma}$. The free energy difference can be obtained by
 integrating the previous equation. In the general case, $U(\gamma)$ is not
 known. The common strategy is to perform several constrained MD
 simulations at different values of $\gamma$ and then integrate
 Eq.(\ref{therm-int}) numerically. Many calculations may be necessary in
 order to integrate Eq.(\ref{therm-int}) with sufficient precision.

 A knowledge of the functional form of the energy would greatly simplify
 this procedure, reducing the number of calculations and allowing the
 analytic integration of Eq.(\ref{therm-int}). The Landau theory in
 combination with MD simulations can provide such useful information. In
 order to apply this formalism, it is necessary to define the thermodynamic
 variables of Eq.(\ref{energy}) from a MD run at finite
 temperature. Statistical mechanics allows us to calculate the order
 parameters by averaging the corresponding time-dependent ones over all the
 available atomic configurations.

 The primitive $c$ cell is unstable with respect to 3 modes of vibration
 whose frequency is degenerate. The instability appears at the $X$, $Y$,
 and $Z$ points of the primitive BZ and the corresponding eigenmodes
 distort the anion sublattice along the $x,y$, and $z$ directions.  In the
 following we consider a supercell which is not the primitive one, and
 those points, originally at the border of the BZ, are folded in at the
 $\Gamma$ point. The eigenvectors are therefore real.

 Let us denote by $\bf u$ the atomic displacements from a perfect site of
 the high symmetry phase. We expand $\bf u$ in normal coordinates using the
 notation of Maradudin {\it et al.}~\cite{Maradudin63}:
\begin{equation}
{\bf u}(\kappa ) = \frac{1}{\sqrt{M_\kappa}}
\sum_{j} {\bf e} ( \kappa | j ) \, Q(j) .
\label{uak}
\end{equation}
 $\kappa$ and $M_\kappa$ label the atoms and their mass in the cell, ${\bf
 e}( \kappa | j )$ is the eigenvector $j$ at the $\Gamma$ point of the BZ,
 and $Q(j)$ is the corresponding normal coordinate. We will denote by
 $|\alpha)$, where $\alpha=x,y,z$, the indexes $j$ describing the soft
 modes.

 Given a general atomic configuration at time $t$ (we now include the time
 in the notation), we define the {\em time-dependent} order parameter as
 the average displacement along $X_2^-$ of the $r_{\rm O}$ oxygen atoms of
 the cell:
\begin{eqnarray}
\delta_\alpha (t) & \equiv & \frac{1}{\sqrt{r_{\rm O}}} \sum_\kappa {\bf u} 
(\kappa,t) \cdot {\bf e}(\kappa | \alpha) \label{def-orp}  \\
& \equiv & \frac{Q (\alpha,t ) }{\sqrt{r_{\rm O} M_{\rm O}}} 
 . \nonumber 
\end{eqnarray}
 The time averages of these quantities, $\bar{\delta}_\alpha$, are the
 experimentally measurable order parameters which we now take as
 thermodynamic variables.

 The factor $\partial U(t) /\partial \delta_\alpha$ entering in
 Eq.(\ref{therm-int}) can now be calculated at each time step by applying
 the definition of $\delta$ given in Eq.(\ref{def-orp}), and by using the
 chain rule:
\begin{equation}
\frac{\partial U(t)}{\partial \delta_\alpha (t)} = 
 \sqrt{r_{\rm O} M_{\rm O}} \, \, 
\sum_{\kappa } \left( \frac{\partial U(t)}{\partial {\bf u}
(\kappa,t)} \right) \cdot  \frac{\partial {\bf u} (\kappa,t)} 
{\partial Q(\alpha,t)} .
\end{equation}
 Noting that the eigenvectors are orthonormal and that the first term of
 the sum is the force ${\bf F}$ acting on the atoms we end up with the
 following expression:
\begin{equation}
\frac{\partial U(t)}{\partial \delta_\alpha (t)} = \sqrt{r_{\rm O}} \,
\sum_\kappa - {\bf F} (\kappa,t) \cdot {\bf e} ( \kappa |
\alpha )  \, . \label{partial-Udlt}
\end{equation}
 Therefore the free energy gradient is calculated from the time average of
 the atomic forces projected along the $X_2^-$ mode of vibration:
\begin{equation}
\frac{d F}{d \delta_\alpha } = \sqrt{r_{\rm O}} \,
\left\langle \sum_\kappa - {\bf F} (\kappa,t) \cdot {\bf e}
( \kappa | \alpha )  \right\rangle_{\bar{\delta}_\alpha} \, . \label{free-grad}
\end{equation}
 Note that the above average has to be taken on an ensemble with a
 constant value of order parameter $\bar{\delta}_\alpha$, {\it i.e.} it
 is necessary to constrain the order parameters during the MD
 simulations.

 \subsection{Constraining the order parameters} \label{constr-orp}

 The dynamics of canonical and microcanonical {\it ensembles} with fixed
 cell shape automatically constrain the auxiliary order parameters. On
 the contrary, in order to constrain the dynamics of the primary order
 parameters and then integrate Eq.(\ref{free-grad}) from the results of
 the MD simulation, it is necessary to modify the Lagrangian of the
 system.~\cite{Frenkel96} 

 The goal is to obtain an equation of motion describing the time evolution
 of a system with a fixed order parameter $\bar{\delta}$. This is done by
 extending the Lagrangian of the unconstrained system ${\cal L}^u$:

\begin{equation}
{\cal L} = {\cal L}^u - \sum_\alpha \lambda_\alpha \sigma_\alpha .
\label{ext-lagr}
\end{equation}
 The superscript $u$ stands for unconstrained, the $\lambda$'s are the
 Lagrange multipliers to be calculated and the $\sigma$'s are the functions
 describing the constraints. Three of them are needed, one for each
 direction $\alpha$ of the tetragonal distortion:
\begin{equation}
\sigma_\alpha(t) = \delta_\alpha(t) - \bar{\delta}_\alpha = 0. \label{constr}
\end{equation}
The Lagrangian of the constrained system is obtained from equations
 (\ref{ext-lagr}) and (\ref{constr}), and the corresponding equations of
 motion are:
\begin{equation}
M_{\rm O} \ddot{\bf u}_\alpha (\kappa,t) = {\bf F}_\alpha (\kappa,t) -
\frac{\lambda_\alpha(t)}{\sqrt{r_{\rm O}}} \, {\bf e}(\kappa | \alpha) .
\end{equation}
 Where ${\bf u}_\alpha$ and ${\bf F}_\alpha$ are the $\alpha=x,y,z$
 components of the displacement and of the force. The orthonormality of the
 normal modes of vibration decouples the equations along the three
 crystallographic directions, simplifying the implementation of the
 method. Moreover, since the tetragonal distortion involves the anion
 sublattice only, we need apply the above modified equation only to the
 $r_{\rm O}$ oxygen atoms. In general, the Lagrange multipliers have to be
 found numerically, but in the present case (decoupled crystallographic
 directions and linear constraints) an analytical solution does exist.

 The expression of the Lagrange multipliers may be found as follow: (i)
 Advance the atomic positions with a {\it fake} unconstrained MD step.
 (ii) Use these unconstrained coordinates to find the multipliers which
 exactly satisfy the constraints. (iii) Use these values of $\lambda$'s to
 perform the {\it true} MD step which satisfies the constraining equations
 by construction. Here we specify this procedure for the leapfrog Verlet
 algorithm.
 
 Given a set of atomic positions ${\bf u}(t)$ which satisfy the
 constraining equations (\ref{constr}), the {\it fake} step involves
 solving the equation of motion corresponding to the Lagrangian ${\cal
 L}^u$. By doing so, the set of unconstrained coordinates ${\bf
 u}^u(t+\Delta t)$ is obtained. These are related to the constrained
 atomic positions ${\bf u}(t+\Delta t)$ as follows:
\begin{equation}
{\bf u}_\alpha(\kappa,t+\Delta t) = {\bf u}_\alpha^u(\kappa,t+\Delta t) - 
\frac{ \lambda_\alpha(t) \, \Delta t^2}{\sqrt{r_{\rm O}} \, M_{\rm O} } \, {\bf e}(
\kappa | \alpha) . \label{disp-cstr}
\end{equation}
 Applying the definition (\ref{def-orp}) to these coordinates, a similar
 relationship may be found for the order parameters.
\begin{equation}
\delta_\alpha(t+\Delta t) = \delta_\alpha^u (t+\Delta t) -
\frac{\lambda_\alpha(t) \, \Delta t^2}{r_{\rm O} \, M_{\rm O}}
\end{equation}
 The analytic solution of the Lagrange multipliers is obtained by
 imposing the constraining equations $\sigma(t+\Delta t)=0$ and then
 solving the resulting linear equation in $\lambda$: 
\begin{equation}
\lambda_\alpha(t) = \frac{r_{\rm O} \, M_{\rm O}}{\Delta t^2} \left[
 \delta_\alpha^u (t + \Delta t) - \bar{\delta}_\alpha \right]. \label{lambda}
\end{equation}
 The substitution of (\ref{lambda}) in (\ref{disp-cstr}) gives the
 constrained coordinates at $t + \Delta t$ in the {\it NVE ensemble}:
\begin{eqnarray}
{\bf u}_\alpha(\kappa,t+\Delta t) & = & {\bf u}_\alpha^u(\kappa,t+\Delta
t) \\
 & & - \sqrt{r_{\rm O} }\left[ \delta_\alpha^u(t + \Delta t) - \bar{\delta}_\alpha \right]
 {\bf e}(\kappa | \alpha) . \nonumber
\end{eqnarray}

 It is important to notice that using this method, the expressions for
 the multipliers are functions of both the integrating scheme and any
 other additional constraints, such as thermostats. The simple case of
 the leapfrog Verlet algorithm described here has to be slightly modified
 in order to include the Nos\'e-Hoover
 thermostat.~\cite{Nose84a,Nose84b,Hoover85} The same procedure may be
 repeated for the {\it NVT ensemble} and the resulting equations of
 motion are:
\begin{eqnarray}
\ddot{\bf u}_\alpha(\kappa,t) & = & \frac{{\bf F}_\alpha (\kappa,t)}{M_O} \\
 & & - \sqrt{r_{\rm O}} \left[ \frac{ \delta_\alpha(t + \Delta t) -
\bar{\delta}_\alpha}{\Delta t^2} \left( 1 + \xi(t) \frac{\Delta t}{2}
\right) \right] {\bf e}(\kappa | \alpha), \nonumber
\end{eqnarray}
 where $\xi$ is the thermostat variable. 

 \subsection{Fluctuations} \label{fluct}

 The fluctuations of the instantaneous order parameter $\delta_\alpha(t)$
 were used to calculate the frequency of a particular vibration directly
 from the MD run. The central point of this analysis is the calculation of
 the fluctuation correlation function spectrum:
\begin{equation}
S_\alpha(\nu)=\int e^{ -i 2 \pi \nu t} \left\langle
\delta_\alpha(t=0) \, \delta_\alpha(t) \right\rangle dt. \label{ft}
\end{equation}
 The above dynamic form factor is known to exhibit two important
 features,~\cite{Schneider73} a temperature-dependent resonant peak at
 $\bar{\nu}$, and an additional central peak at $\nu=0$. The relative
 magnitude of the two peaks depends on the transformation mechanism and on
 the temperature. This has been proved for phase transition mechanisms as
 different as order-disorder and displacive.~\cite{Padlewski92} Therefore,
 without loss of generality, following Padlewski {\it et
 al.}~\cite{Padlewski92}, the power spectrum (\ref{ft}) can be modelled as a
 superposition of two peaks with the following functional form:
\begin{equation}
S_\alpha(\nu)= \frac{2 A B}{B^2 + \nu^2} + \frac{ C D}{D^2 + \left( \nu -
\bar{\nu}_\alpha \right)^2}, \label{ft2}
\end{equation}
 where $A$, $B$, $C$, $D$ and $\bar{\nu}_\alpha$ are parameters to be
 fitted to the calculations. The analytical form of the time-dependent
 correlation function $S(t)=\left\langle \delta_\alpha(t=0) \,
 \delta_\alpha(t) \right\rangle$ may be found by substituting (\ref{ft2})
 in (\ref{ft}) and passing into the time domain with an inverse Fourier
 transform:
\begin{equation}
 S_\alpha(t)= A \, e^{-B \, t} + C \, e^{-D \, t} \cos(2\pi
\bar{\nu}_\alpha t). \label{fl-corr}
\end{equation}

 The time-dependent order parameter is calculated from the MD atomic
 positions. The time correlation function of $\delta_\alpha(t)$ is then
 obtained using the multiple time-origin method~\cite{Allen87} and fitted
 to Eq.(\ref{fl-corr}). The fitting procedure provides both the time
 correlation function and its Fourier transform.

 \section{MD Simulations} \label{SecRslt}

 The polarizable self-consistent  Tight-Binding
 model~\cite{strcZrO2,Finnis} was used to perform two sets of
 MD simulations. In the first one, standard MD calculations were used to
 investigate the temperature dependence of the order parameters and to
 follow the softening of the $X_2^-$ mode of vibration up to the
 transition point. In the second one, we combined the constrained MD
 simulations and the thermodynamic integration technique to calculate the
 free energy of the $c \leftrightarrow t$ phase transition, to study the
 nature of the order parameter fluctuations and to explain the high
 temperature stability of the $c$ phase.

 \subsection{Standard MD simulations} \label{SecMDrslt}

 \subsubsection{Softening of a vibrational frequency}

 The time evolution of a system of 96 particles with periodic boundary
 conditions has been followed at temperatures between 300 and 2200 K. The
 lattice parameter of the simulation cell were the correspondent
 experimental values of Aldebert and Traverse, which, where necessary,
 have been linearly extrapolated at lower temperatures. During each MD
 run, the temperature has been constrained with a Nos\'e-Hoover
 thermostat,~\cite{Nose84a,Nose84b,Hoover85} and the equations of motion
 have been integrated for not less then 5 ps with a typical time step of
 5 fs. Near the transition point, the time step has been reduced to 2.5
 fs and the total simulation time has been increased to 15 ps.

 The cell size was constrained by the relatively high number of MD runs
 necessary to follow the phase transition. In total, we simulated the
 time evolution of 96 particles for more than 120 ps. As discussed later,
 the cell size does not change our qualitative description of the phase
 transition, and the 324-atom unit cell would have just implied a heavier
 computational effort, without adding further information to the
 physical picture provided by the smaller cell.

 We started the simulations from the crystallographic positions of the
 tetragonal phase and equilibrated the system at the temperature of 300
 K. This temperature is well inside the field of stability of the $m$
 phase, however, during the MD simulations, the system remained in the $t$
 phase because of the existence of an energy barrier between the two
 structures. We calculated the vibrational frequencies of the $t$ structure
 by diagonalising the dynamical matrix at the origin and at the borders of
 the Brillouin Zone (BZ) along the (100), (110), and (111) directions. This
 analysis showed that all the vibrational frequencies are real and that the
 $t$ phase does not spontaneously distort towards the $m$ structure.

\begin{figure}
\caption{Time-dependent order parameters at 700 K: $\delta_x$ and $\delta_y$
oscillate around 0, $\delta_z$ around $\bar{\delta}_z$, the value of the
macroscopic order parameter.}
\label{orp}
\centerline{\psfig{file=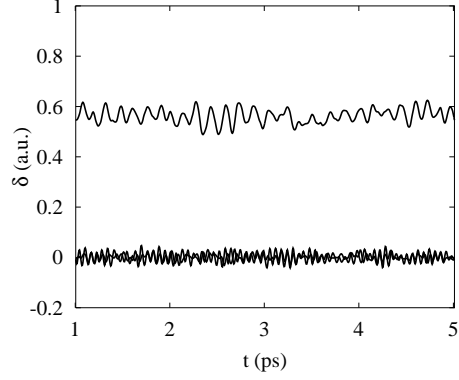,width=6cm,angle=-90}} 
\end{figure}

\begin{figure}
\caption{Time correlation functions (top) and corresponding Fourier
transforms (bottom) of the time-dependent order parameters $\delta_x$, 
$\delta_y$, and $\delta_z$.}
\label{cf-ft}
\centerline{\psfig{file=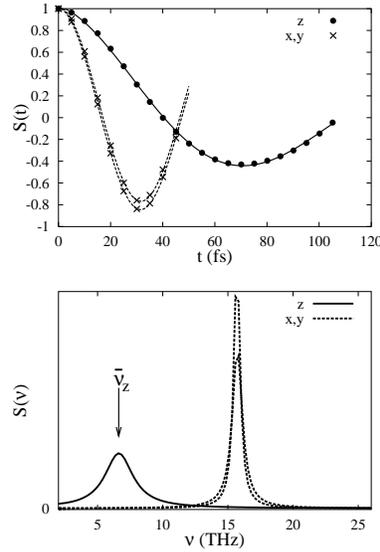,width=5cm,angle=0}} 
\end{figure}

 In this set of MD simulations we followed the approach of Padlewski {\it
 et al.}~\cite{Padlewski92} described in Section~\ref{fluct}, focusing on
 the instantaneous order parameters $\delta_\alpha(t)$ [see
 Eq. (\ref{def-orp})] which fluctuate about the mean value
 $\bar{\delta}_\alpha$. Figure~\ref{orp} shows the typical time evolution
 of the primary order parameters for the MD run at T=700 K.
 Figure~\ref{cf-ft} shows the fluctuation autocorrelation function $S(\nu)$
 and the corresponding frequency spectrum $S(t)$ for the MD run at 700 K:
 the arrow points at $\bar{\nu}_z$, the frequency which softens. It can be
 seen that the $x$ and $y$ components, corresponding to the
 transverse optical frequencies, are degenerate.

\begin{figure}
\caption{Temperature dependence of the macroscopic order parameter
$\bar{\delta}_z$. The symbols ($\bullet$) are the results of the
calculations. The continuous solid eye-line is extrapolated in the region
near $T_c$ where the large fluctuations in $\delta_z$ make the averaging
procedure inaccurate.}
\label{soft3}
\centerline{\psfig{file=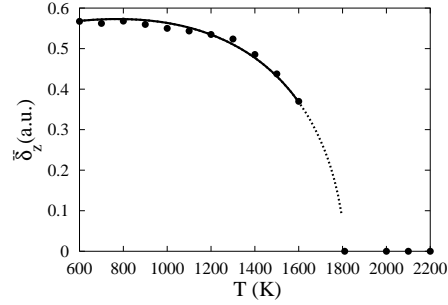,width=6cm,angle=0}} 
\end{figure}

\begin{figure}
\caption{Calculation results of the frequency squared $\bar{\nu}_z^2$
vs. temperature for two simulation sets : (a) tetragonal cell with
temperature-dependent lattice parameters taken from
experiment,~\cite{Aldebert85} (b) Cubic cell with temperature-independent
lattice parameter (see text).}
\label{soft}
\centerline{\psfig{file=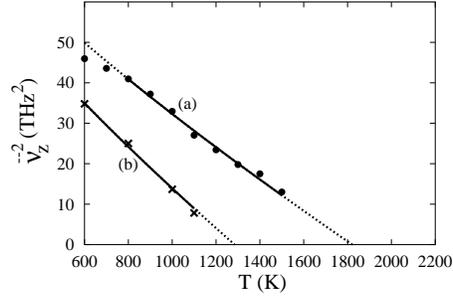,width=6cm,angle=0}} 
\end{figure}

 On increasing the temperature, the softening of the frequency
 $\bar{\nu}_z$ is evident from the dynamic form factor, where the resonant
 peak shifts. At the same time, the primary order parameter decreases
 continuously (Figure~\ref{soft3}), as experimentally observed in the
 similar system ZrO$_2$-12\%ErO$_{1.5}$.~\cite{Yashima93} The calculated
 temperature dependence of the macroscopic order parameter $\bar{\delta}_z$
 and of the corresponding vibrational frequency, shown in
 Figure~\ref{soft}, was then interpreted using the Landau theory.
 We found that the critical exponent for this phase transition is
 $\beta$=0.35. According to the same theory, the critical exponent
 $\beta^\prime$ for the auxiliary order parameters ($\eta_2,\eta_3$)
 describing the tetragonality of the cell is bigger then $\beta$. Therefore
 the $c/a$ ratio should depend more strongly on the temperature than $\delta$.

 As the transition temperature is approached, the decrease of the order
 parameter and of the corresponding frequency is accompanied by an
 increase in the order parameter fluctuations, which theoretically
 diverge at $T_c$ for a second order phase transition. As a result, it
 was not possible to follow the complete softening of the frequency:
 there is a temperature window about $T_c$ where, even though long MD
 simulations allow one to evaluate the average order parameters, it is not
 possible to calculate the frequency. In this temperature range, the
 frequency $\bar{\nu}$ is so low that the corresponding peak in the
 dynamic form factor $S(\nu)$ merges with the central peak and it is not
 possible to separate them. 

 The theoretical transition temperature of $\approx$1800 K is $\approx$30\%
 lower than the experimental value of $\approx$2600.~\cite{Aldebert85} This
 may be explained by noting that the first-principles calculations
 underestimate the energy difference between the $c$ and $t$ structures
 $\Delta U^{t-c}$, which determines
 $T_c$.\cite{strcZrO2,Kralik98,Stapper99} This is the ab initio energy
 barrier between the minima of the double well, which was used to
 parametrize the SC-TB model. In particular the SC-TB results underestimate
 the experimental $\Delta U^{t-c}$ by 30\%, which is consistent with the
 underestimate of the transition temperature.

 According to the renormalized phonon group theory,~\cite{Bruce80}
 $\bar{\nu}^2$ depends linearly on the temperature in both the regions
 $T<T_c$ and $T>T_c$, and the correspondent slopes are related by the
 following relationship:
\begin{equation}
R=\left(\frac{d \bar{\nu}^2}{dT}\right)_{T<T_c} /
  \left(\frac{d \bar{\nu}^2}{dT}\right)_{T>T_c}=-2. \label{R}
\end{equation}
 However, our simulations at $T>T_c$ suggest that the $c
 \leftrightarrow t$ phase transition in zirconia has a different
 behaviour from the ideal case described by Eq.(\ref{R}), because no
 frequency was observed above $T_c$.

 The exploration of the high temperature region of the zirconia phase
 diagram has been carried out in two stages. As a first attempt, we
 continued the MD simulations on the system described above, simply
 increasing the temperature. This has been done up to 2200 K. The time
 autocorrelation function (\ref{fl-corr}) of these simulations
 exponentially decayed without showing any structure. As a result, the
 central peak dominated the corresponding dynamic form factor, and
 therefore it was not possible to isolate the resonant peak at $\bar{\nu}$
 from the central one. A possible explanation of this may be proposed by
 noticing that, according to Eq.(\ref{R}), for $T>T_c$ the slope of
 $(\frac{d \bar{\nu}^2}{dT})$ is half that for $T<T_c$. This means that the
 temperature window around $T_c$, in which it is not possible to calculate
 the frequency, extends more in the high temperature field than in the low
 temperature one. Probably 2200 K is still in the region of {\it
 disturbance} of the transition point. 

 In order to verify if the frequency does eventually increase in the high
 temperature region, we studied a similar system with the same properties
 of that one described above but with a lower transition temperature. The
 idea is based on the following argument. It is well established that the
 relative energetics of the two phases is governed by a double well in
 the potential energy that depends on volume and
 $c/a$.~\cite{Jansen88,Jansen91,Orlando92,Wilson96zirc} We studied in
 detail its dependence,~\cite{strcZrO2} which is also captured by the
 Landau expansion (\ref{energy}). Both the hydrostatic and tetragonal
 strains modify the double well in the same way: the smaller the volume
 (or the $c/a$ ratio), the smaller the energy difference and therefore
 the smaller the transition temperature. Incidentally, this is connected
 to the ab initio underestimate of the energy barrier, which is
 calculated with the structural parameters corresponding to 0 K.

 By exploiting this property of the energy surfaces, we made a new set
 of MD simulations aimed to explore the temperature range $T>T_c$. In
 these calculations the volume was chosen to lower the transition
 temperature to $\approx$1300 K and the cell was kept cubic ($c/a=1$)
 even in the low temperature region. As expected, for $T<T_c$, the linear
 softening of $\bar{\nu}^2$ (Figure~\ref{soft} set b) was obtained for
 this system as well. The slope was slightly different because in the
 previous simulations the thermal expansion of the cell was included in
 the description, while in this case the volume was fixed to the initial
 value. Because of this we could calculate the frequency up to within
 $\approx$200 K of the transition temperature. The temperature was then
 increased to 2000 K. Surprisingly, even in this case, there was no
 structure in the autocorrelation function $S(t)$ and the expected {\it
 hardening} of the frequency was not observed. This suggests that in the
 $c$ phase the motion of the oxygen sublattice along the $X_2^-$ mode of
 vibration is, in terms of $S(t)$, uncorrelated. This behaviour will be
 clarified by the free energy surfaces described in
 Section~\ref{SecFreeSurf}.

 \subsection{MD simulations at constant {\boldmath  $\delta$}} 
   \label{SecFreeSurf}

 In our previous papers~\cite{strcZrO2,Finnis} we restricted the analysis
 of the 0 K energy surface to one tetragonal invariant only. By doing so,
 we defined a simplified version of the energy expansion (\ref{energy})
 involving the strain and one component of the primary order parameter. We
 then fitted the correspondent coefficients, $a_2$, $a_{41}$, $b_1$, $b_2$,
 $c_1$, and $c_2$, to the results of total energy calculations. We also
 showed how the coupling between the primary and auxiliary order parameters
 could create a critical point where the transformation becomes first
 order. Here that analysis is extended by exploring the topology of the
 energy surface in the whole $\delta$ domain, and by following its
 temperature evolution through the phase transition into the field of
 stability of the $c$ phase. This sheds light on the mechanism of the phase
 transformation and on the high-temperature stability of the $c$ phase.

 The following results were obtained using a 12-atom unit cell with
 different $c/a$ ratios (1, 1.01, 1.02) at the 0 K theoretical equilibrium
 volume of the $c$ structure. Preliminary unconstrained MD simulations were
 done to explore the effect of the cell size on the physical picture of the
 phase transformation described in the previous Section. Even in this small
 system, the frequency $\bar{\nu}_z^2$ depends linearly on the temperature
 and the predicted transition temperature is of $\approx 1600$ K. The
 effect of using a small cell is to shift $T_c$ to higher
 temperatures. This is consistent with the physical picture proposed in
 Section~\ref{SecMDrslt}: the autocorrelation function $S(t)$ measures the
 degree of correlation between the motion of the oxygen atoms along the
 $X_2^-$ mode of vibration. We described how the temperature acts on $S(t)$
 by reducing the correlation until this is completely lost above $T_c$,
 where the corresponding frequency is soft, and where the structure is
 $c$. The small cell size and the periodic boundary conditions force the
 motion of atoms in adjacent cells to be correlated, and therefore
 counteract the effect of the temperature on $S(t)$. As a result, in the
 small system, higher temperatures are needed to observe the complete
 softening of the frequency $\bar{\nu}_z$.

 The 12-atom and 96-atom supercells have the same temperature
 dependence of $\delta_z$ and $\bar{\nu}_z$ but the corresponding curves
 are shifted to different temperatures. We can therefore conclude that
 the phase transformation mechanism is the same in the two systems. We
 shall calculate the
 free energy of the transition from the MD simulations
 of the small cell, and assume that the resulting qualitative physical
 picture applies to bigger cell sizes.

 Before exploring the free energy temperature dependence, it is
 useful to simplify the complete energy expansion (\ref{energy}) by
 neglecting the order parameters which are unlikely to play an important
 role in the phase transition. The transformation between the $c$ and $t$
 structures does not distort the cell shape as described by the order
 parameters ($\eta_4,\eta_5,\eta_6$). It is therefore reasonable to
 neglect them in the discussion of the following results. Moreover, even
 though the transformation between the $c$ and the $t$ structure does
 involve a change in the volume, the energetic contribution of the
 associated order parameter $\eta_1$ is well understood and has already
 been discussed. Apart from the 0 K case, we will not consider the terms
 $B_3$ and $C_3$ in the energy expansion. However, their possible
 influence on the character of the phase transition in terms of softening
 of the corresponding elastic constant, will be discussed {\it a
 posteriori} in the final Section~\ref{Sec-ct2}.

 \subsubsection{Topology of the 0 K surface}

\begin{figure}
\caption{Section of the 0 K energy surface for the cubic cell. The
isoenergetic contours are plotted on the base every 0.3 mRy/ZrO$_2$.}
\label{surf-t0}
\centerline{\psfig{file=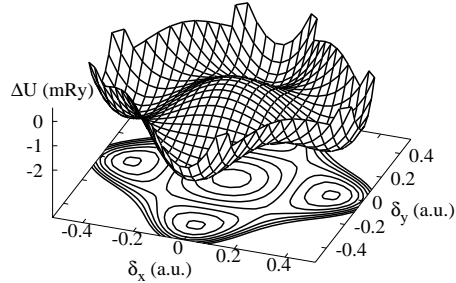,width=6cm,angle=-90}} 
\end{figure}

 We start our analysis with the primary order parameter. Two sets of
 calculations on a stress-free cubic unit cell were used to fit the
 coefficients $a_2$, $a_{41}$, and $a_{42}$. These have been determined by
 distorting the oxygen sublattice along $\langle \delta 00 \rangle$ and
 along $\langle \delta \delta 0\rangle$. We plot the resulting energy
 surface, which we take as starting point of our analysis, as a function of
 two tetragonal invariants in Figure~\ref{surf-t0}. In this simple case,
 because of the cubic cell, the three components of the primary order
 parameter are equivalent.

\begin{figure}
\caption{0 K energy surfaces for a tetragonal cell with the tetragonal
axis along $z$: (a) section in the $\delta_y,\delta_x$ plane, (b) section
in the $\delta_z,\delta_x$ plane. The isoenergetic contours are plotted
on the base every 0.3 mRy/ZrO$_2$.}
\label{surf-t0tet}
\centerline{\psfig{file=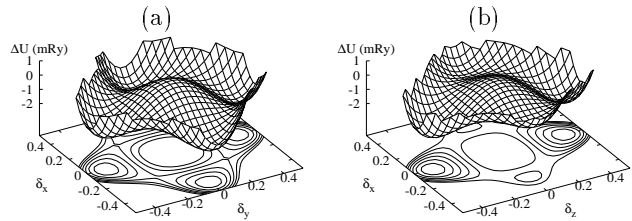,width=8.5cm,angle=0}} 
\end{figure}

\begin{figure}
\caption{Transferability of the 0 K energy-expansion coefficients between
different tetragonal cells with the tetragonal axis along
$z$. Projections of the corresponding energy surfaces along the
high-symmetry order-parameter directions $\langle 00\delta \rangle$ (a), and
$\langle \delta \delta 0 \rangle$ (b).}
\label{lndchk}
\centerline{\psfig{file=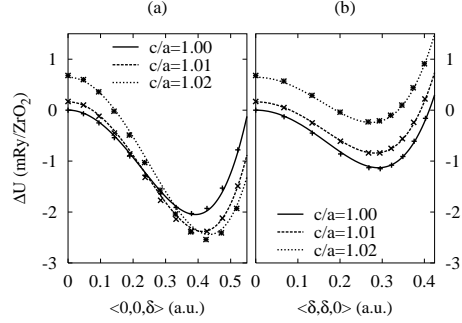,width=6cm,angle=-90}} 
\end{figure}

 The same set of calculations was then done on a tetragonal cell
 ($c/a=1.01$), by which we determined the parameters $b_2$ and $c_2$. The
 latter is proportional to the elastic constant $C^\prime$. The
 transferability of the parameters was then checked by redoing the
 calculations for a different tetragonal cell ($c/a=1.02$):
 Figure~\ref{lndchk} shows that the same set of coefficients fit the
 results for this cell as well. If $z$ is the tetragonal axis, the
 tetragonality of the unit cell shortens the average interatomic distances
 in the transverse $x,y$ plane and lengthens them along the tetragonal
 axis. As a consequence, the energy surface section in the transverse plane
 $\delta_x,\delta_y$, shown in Figure~\ref{surf-t0tet} (a), is similar to
 the reference one of Figure~\ref{surf-t0} but shallower and tighter, while
 it is deeper and broader along $\delta_z$ [Figure~\ref{surf-t0tet} (b)].
 
 Finally, the same procedure described above was used to fit the remaining
 coefficients $b_1$ and $c_1$ by distorting the cell with respect to the
 order parameter $\eta_1$ defined in Table~\ref{op-irrep}.

 In conclusion, the static calculations show that the 0 K energy surfaces
 can be captured by Taylor expansion up to fourth-order and
 are therefore completely defined by the set of coefficients given in
 Table~\ref{coeff}.

\subsubsection{Free energy surfaces}

 The MD simulations were carried out in the temperature range from 50 K
 to 2000 K, constraining the primary and secondary order parameters. Let
 us first focus on the results for the cubic cell, commenting later on the
 effect of the $c/a$ ratio. The explorations along the directions
 $\langle \delta 00 \rangle$ and $\langle \delta  \delta 0
 \rangle$ fully determine the free energy surfaces to fourth order,
 therefore we constrained the order parameters along these directions
 from 0 to 0.7 a.u., using the dynamics described in
 Section~\ref{constr-orp}. The quantity defined in
 Eq.(\ref{partial-Udlt}) was accumulated during the MD run and its time
 average provided the ensemble average required in
 Eq.(\ref{free-grad}). The analytical form (\ref{energy}) of the energy
 surface was then differentiated along the corresponding direction and
 fitted to the results of the simulations. For this particular case, the
 fit provided both the free energy gradient and the free energy
 itself. This is because we chose the reference energy as the top of the
 double well for a cubic crystal. The integration of Eq.(\ref{free-grad})
 provides the energy difference $\Delta F$:
\begin{equation}
\Delta
F=F(\epsilon,\bar{\delta})-F(\epsilon,\bar{\delta}=0)=\int_0^{\bar{\delta}}
\left\langle \frac{\partial U}{\partial \delta}
\right\rangle_{\bar{\delta}} d\delta. \label{deltaF}
\end{equation}
 We arbitrarily set to zero the integration constant for the cubic cell
 $F(\epsilon=0,\bar{\delta}=0)$. If the cell is tetragonal the same
 constant is $c_{2} \, C_{2}(\epsilon^2) /2$. Therefore, the fit of the
 free energy gradient to the MD simulations of the cubic cell provided
 the coefficients $a_2$, $a_{41}$, $a_{42}$ and $b_{2}$ at the
 corresponding temperature.

\begin{figure}
\caption{Temperature-dependent free energy gradients calculated using
Eq.(\ref{free-grad}) and corresponding fit via the analytic form derived
from the Landau theory (\ref{energy}). Projection along the order
parameter direction $\langle 0 0 \delta \rangle$.}
\label{ffree-grad}
\centerline{\psfig{file=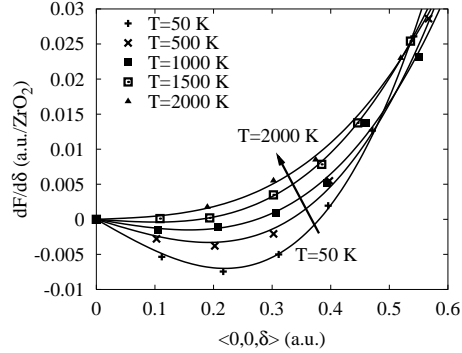,width=6cm,angle=-90}} 
\end{figure}

\begin{figure}
\caption{Temperature evolution of the free energy profiles projected
along the order parameter direction $\langle 0 0 \delta \rangle$. The symbols
($\bullet$) are the 0 K calculations, the thick solid lines are the
result of the thermodynamic integration and correspond to the
temperatures 50, 500, 1000, 1500, and 2000 K.}
\label{well-free}
\centerline{\psfig{file=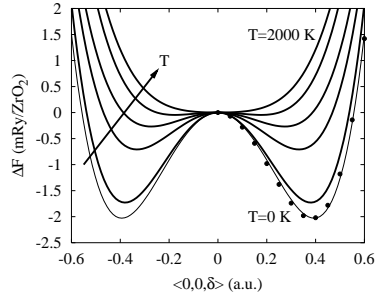,width=6cm,angle=-90}} 
\end{figure}

 The fit to the computed results along the $\langle \delta 00 \rangle$
 and the corresponding free energy profiles obtained from
 Eq.(\ref{deltaF}) are shown in Figures~\ref{ffree-grad} and
 \ref{well-free} respectively. The corresponding expansion coefficients
 are included in Table~\ref{coeff}. Also at high temperature the fourth
 order energy expansion well describes the calculated free energy
 gradients. The temperature acts on the double well by gradually reducing
 the energy difference between the distorted and undistorted
 structure. As expected, the energy surface is very flat near the
 critical temperature, but, surprisingly, it remains quite flat even at
 higher temperatures, well inside the field of stability of the cubic
 phase. The energy surfaces above the transition temperature are highly
 anharmonic.
 
 The unconstrained MD simulations described in the previous Section
 generated the temperature dependence of the $X_2^-$ vibration
 frequency (Figure~\ref{soft}) up to the transition point only. Above the
 critical temperature it was impossible to calculate the frequency
 $\bar{\nu}_z$, the dynamic form factor $S(\nu)$ being dominated by a wide
 central peak. The soft frequency $\bar{\nu}_z$, together with the large
 fluctuations of the order parameters $\delta_z$, suggests a disordered
 dynamics of the oxygen sublattice along the $X_2^-$ mode of vibration. The
 anharmonicity of the energy surfaces explains this behaviour. It may also
 explain the fact that the optical-phonon branches are not experimentally
 observed~\cite{Liu87} in cubic stabilized zirconia.

 As in other perfect fluorite structures,~\cite{Willis75} the
 vibrational motion of the anions in $c$ zirconia appears to be
 anharmonic. This is consistent with the neutron powder diffraction
 experiments of Kisi and Yuxiang~\cite{Kisi98} on cubic stabilized
 zirconia (ZrO$_2$-9.4\%Y$_2$O$_3$). They measured the temperature
 dependence of the Debye-Waller factor and proposed different models to
 fit the data. Both a simple Debye model and a Debye model plus a static
 disorder component provided a poor fit of the data. A radical
 improvement of the fit was obtained when an isotropic anharmonic
 vibration of both species was included in the description.

\begin{figure}
\caption{Double well in the internal energy along the order parameter
direction $\langle 00 \delta \rangle$: the solid line correspond to the 0 K
calculation, the symbols are the averaged internal energies from the
constrained MD simulation.}
\label{int-eng}
\centerline{\psfig{file=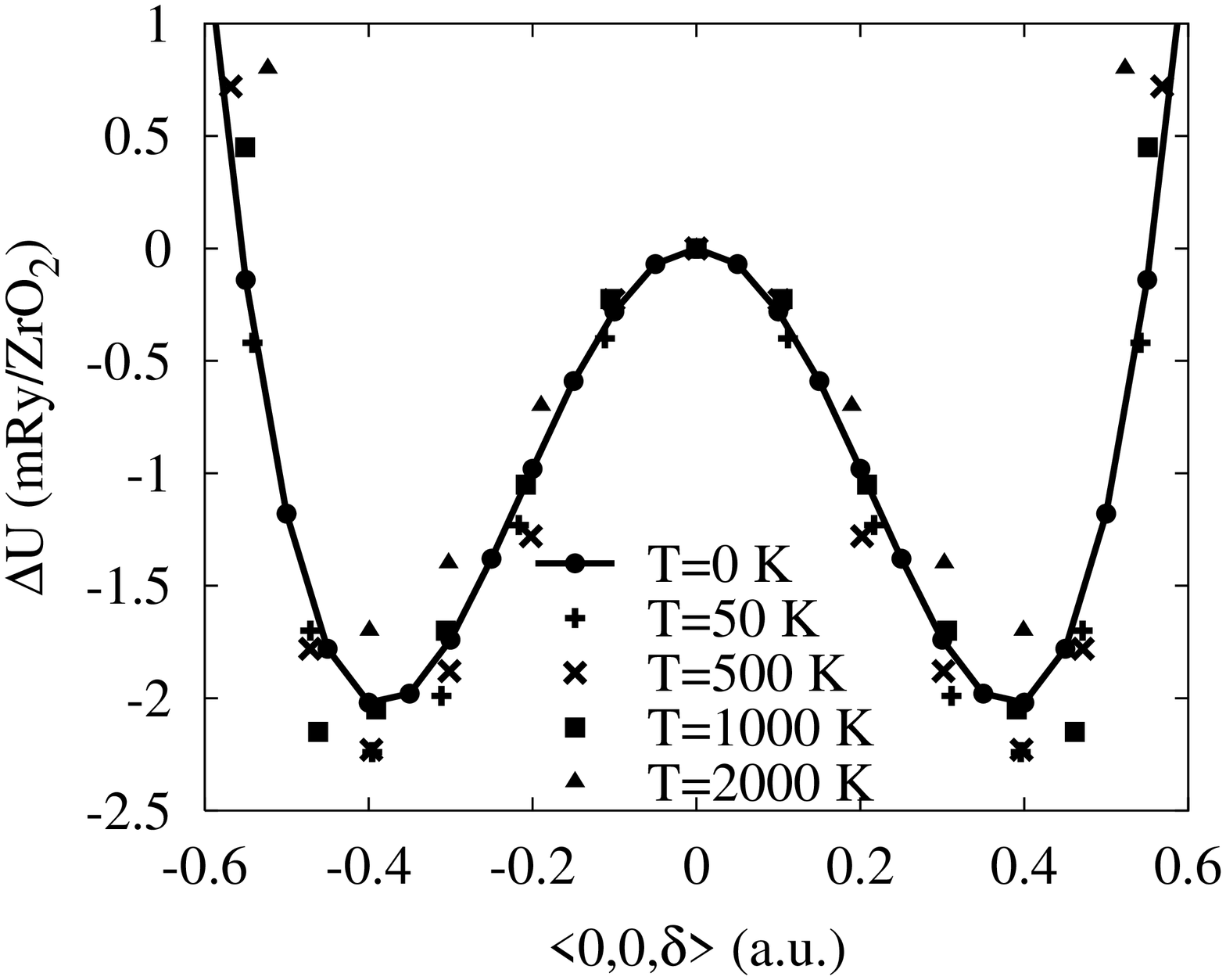,width=6cm,angle=-90}} 
\end{figure}

\begin{figure}
\caption{Entropic contribution to the phase transition obtained by
applying the definition of the Helmholtz
free energy $\Delta F = \Delta U - T \Delta S$ to the data shown in
Figures~\ref{well-free} and~\ref{int-eng}.}
\label{entrpy}
\centerline{\psfig{file=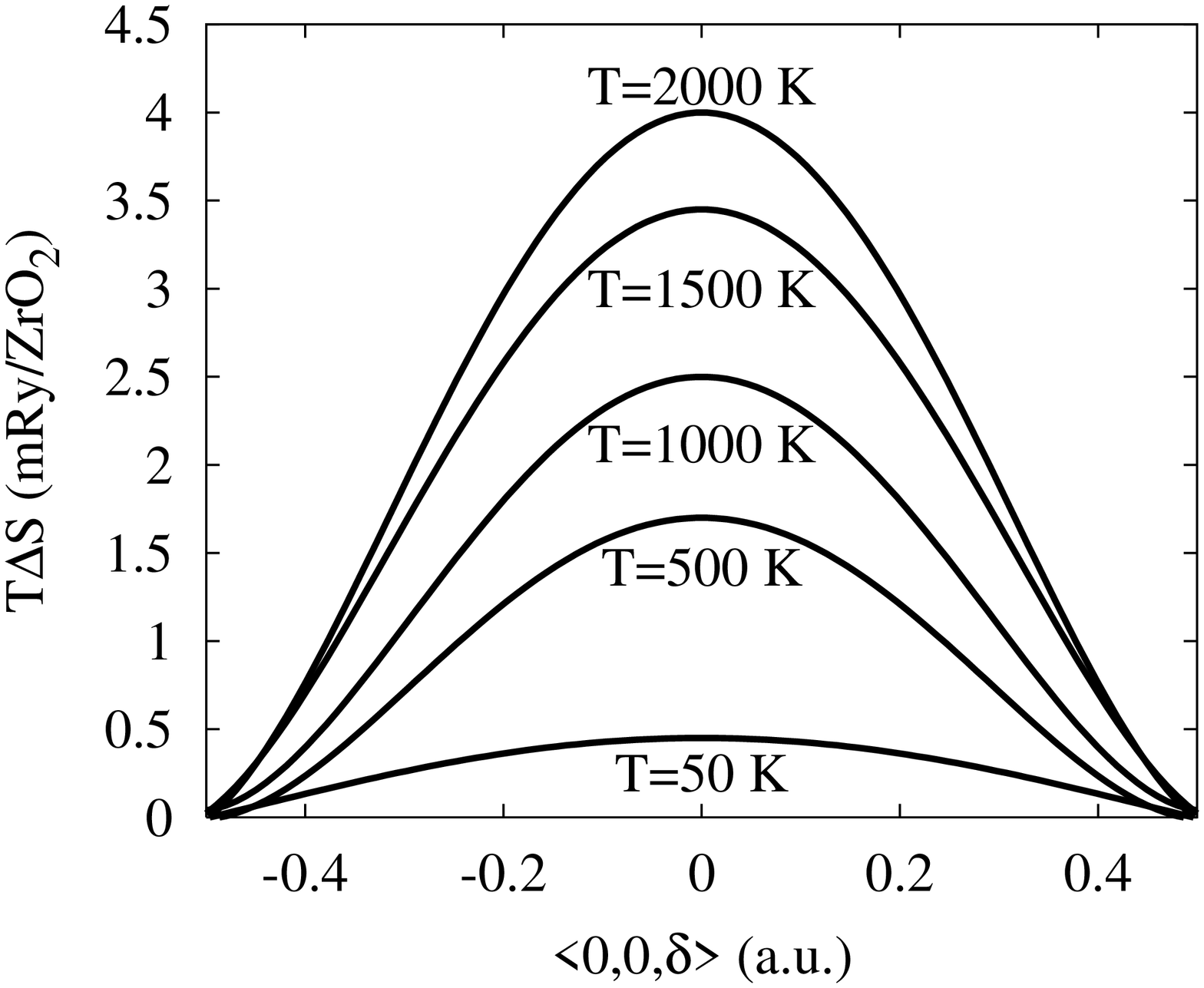,width=6cm,angle=-90}} 
\end{figure}

 We may further investigate the nature of the double well temperature
 dependence by splitting the free energy into its energetic and entropic
 contributions. The time average of the internal energy $U$ from each
 constrained MD simulations is plotted in Figure~\ref{int-eng}. It is clear
 that the double well in the internal energy is present even at $T>T_c$.
 The internal energy double well is relatively insensitive to the temperature
 and the 0 K coefficients of Table~\ref{coeff} provide an excellent fit of
 the finite temperature results for both the low and high temperature
 structures. From the definition of $F$, the difference between the free
 energy calculated with Eq.(\ref{deltaF}) and the internal energy gives the
 entropic contribution $TS$ plotted in Figure~\ref{entrpy}. The high
 temperature stability of the $c$ phase is therefore ensured by this
 entropic term which change the shape of the energy surface from double to
 single welled.

 \subsubsection{Coupling to the elastic strains}
 \label{Sec-ct2}

 The fit of the Landau energy expansion (\ref{energy}) to the calculation
 results has allowed us to follow the temperature evolution of the free
 energy surface through the phase transition. The fitting coefficients for
 each temperature are included in Table~\ref{coeff}. We can see that at the
 critical temperature of $\approx 1600$ K, when $a_2$ goes to zero, the
 fourth order coefficient $a_4$ is positive. Therefore, in this cell, the
 phase transition is diffusionless displacive and second order.

 If the thermodynamic potential $F$ is expanded in terms of elastic
 strains as well as $\delta$, the coupling between the primary order
 parameter and the strain in (\ref{energy}) renormalizes the fourth-order
 coefficient~\cite{Landau5,Cowley80,Anderson65} and could make it small
 or negative near the transition point. As a result, the transformation
 may become first-order.  This could happen if the temperature reduces an
 elastic constant $c_{i}$.

 In order to see this, let us consider a tetragonal cell at the reference
 volume, whose tetragonal axis is $z$ and where the oxygen sublattice is
 distorted along the order parameter direction $\langle 00 \delta_z
 \rangle$. With these restrictions the energy expansion (\ref{energy})
 has a simplified form:
\begin{equation}
F = F_0 + \frac{a_2}{2} \delta_z^2 + \frac{a_{41}}{4} \delta_z^4 + b_2 \,
       \, 2 \delta_z^2 \, \, \eta_2 + \frac{c_2}{2} \, \, \eta_2^2
        + {\cal O}\left (\delta_L^6 \right) \, .\label{energy2}
\end{equation}
 At equilibrium the two order parameters $\delta_z$ and $\eta_2$ are not
 independent and the relationship between the two may be found by imposing
 the equilibrium condition:
\begin{equation}
\frac{\partial F}{ \partial \eta_2} = 0 \hspace{0.5cm} 
\Rightarrow \hspace{0.5cm} 
\eta_2= - \frac{2b_2}{c_2} \delta_z^2. \label{del-eps}
\end{equation}
 A similar procedure may be repeated for the hydrostatic strain
 $\epsilon_{xx}+\epsilon_{yy}+\epsilon_{zz}=\eta_1$ and the substitution of
 Eq.(\ref{del-eps}) back in Eq.(\ref{energy2}), together with the
 corresponding one for $\eta_1$, clarifies the combined effect of the
 coefficients $b_i$ and $c_i$ on the renormalization of the fourth order
 coefficient:
\begin{equation}
F = F_0 + \frac{a_2}{2} \delta_z^2 
  + \left( \frac{a_{41}}{4} - \frac{2 b_2^2}{c_2} -\frac{b_1^2}{2 c_1}
    \right) \delta_z^4 
  + {\cal O}\left (\delta_z^6 \right) \, .\label{renorm}
\end{equation}
 It may be verified that the above formulation is independent of the
 initial choice of the tetragonal axis. 

 The coefficients $c_1$ and $c_2$ are proportional to the bulk modulus
 and to the elastic constant $C^\prime$, respectively. If one or both of
 these quantities significantly decrease with temperature, the
 correspondent term in Eq.(\ref{renorm}) may dominate the sign of the
 fourth-order coefficient, making it very small or negative. At 0 K the
 coupling terms $2 b_2^2/c_2$ and $b_1^2/2 c_1$ reduce the fourth-order
 coefficient by $16$\% and $4$\% respectively. Because of this difference
 we focused our attention on the elastic constant
 $C^\prime$. Experimentally, the high temperature
 data~\cite{Liu87,Liu87b,Kandil82} do not show any anomalous temperature
 dependence of the elastic constants, with a general decrease of
 $\approx15-20$\% between 300 and 1700 K. If this is valid for pure
 zirconia as well, we may anticipate that the degree of softening of the
 elastic constants will not affect the character of the phase transition.

 The calculations described above for the cubic cell have been repeated
 for two tetragonal cells with $c/a=1.01$ and $1.02$. The constrained MD
 simulations along the $\langle \delta 00 \rangle$ direction allowed the
 fit of the coupling coefficient $b_{2}$ to each temperature. Therefore
 it has been possible to obtain the temperature and $c/a$ dependent free
 energy curve along this direction. However, in this case we are
 interested in the relative position of the free energy surfaces
 for the different cells, depending on the integration
 constant $F(\epsilon,\delta=0)=c_{2} \, C_{2}(\epsilon^2) /2 $.  In
 principle one could calculate $c_2$ by a simple thermodynamic
 integration, but this would require monitoring the stress, which is not
 yet implemented in the current program. Therefore we decided to continue
 the analysis by choosing an extreme scenario, which would be a strong
 temperature dependence of $C^\prime$. With a large safety margin with
 respect to the experimental values, we linearly reduced this elastic
 constant by $75$ \% from 200 to 50 MPa between 0 and 2000 K.

\begin{figure}
\caption{Free energy profiles projected along the order parameter
direction $\langle 00 \delta \rangle$ for cubic and tetragonal cells
below and above the transition temperature.}
\label{free-ca}
\centerline{\psfig{file=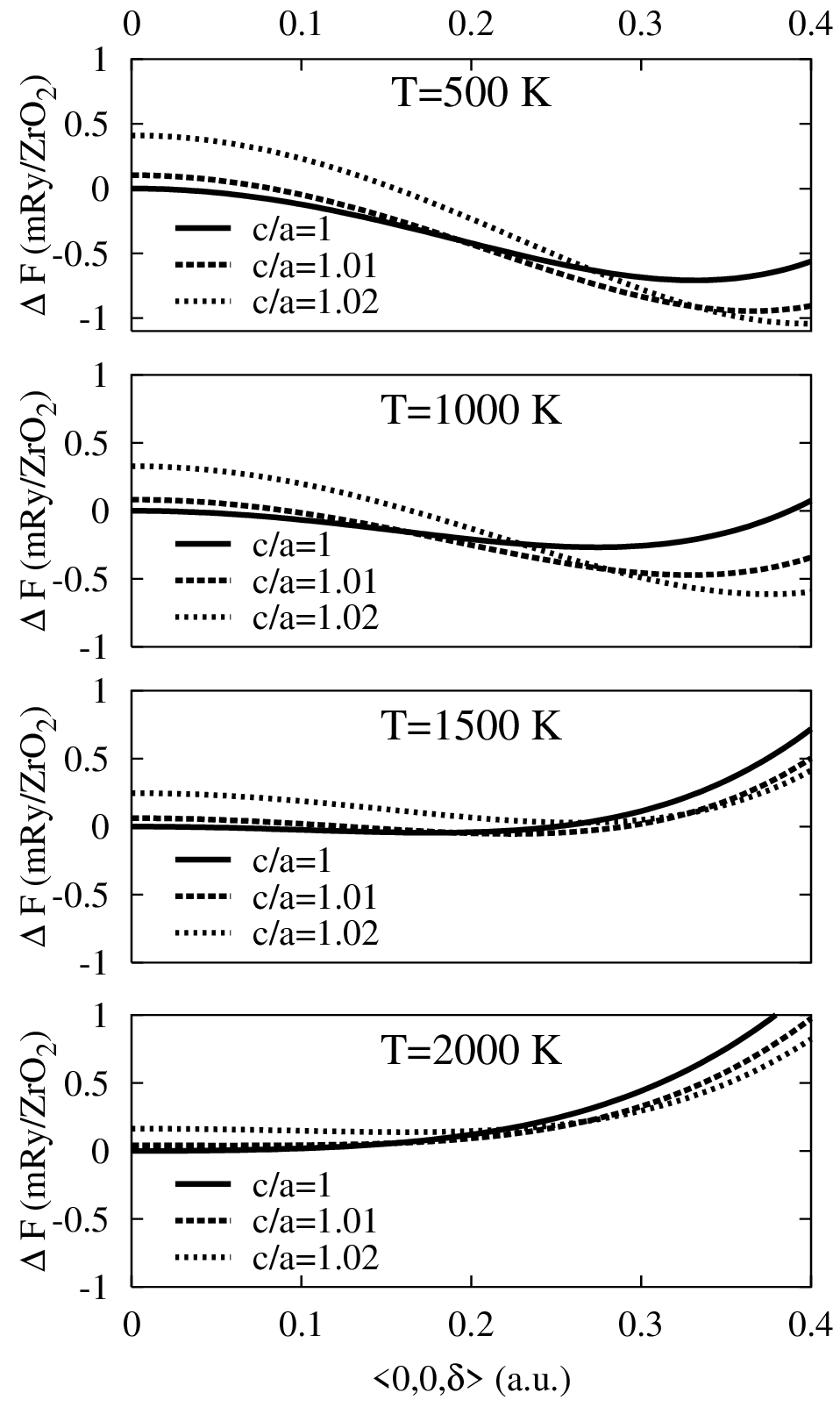,width=6cm,angle=0}} 
\end{figure}

 With this assumption, the sections of the free energy surfaces at
 different temperatures between 500 and 2000 K are plotted in
 Figure~\ref{free-ca} as a function of $\delta$ and $c/a$. As
 experimentally observed,~\cite{Aldebert85} the tetragonal cell with
 $c/a=1.02$ is thermodynamically stable up to very near $T_c$ where the
 minimum configuration goes from $(c/a=1.02,\delta=0.4)$ to
 $(c/a=1,\delta=0)$ quickly but continuously. We believe that the sudden
 change of the order parameters, due to the flat energy surfaces near the
 transition point and therefore to the anharmonicity of the material, may
 explain the fact that this phase transition has been considered to be
 first order in the early studies.

 It is interesting to note that both the $a_{41}$ and the coupling
 coefficient $b_2$ decrease with temperature
 (Table~\ref{coeff}). Because of this, even with the large postulated
 softening of $C^\prime$, the renormalization of the fourth-order
 coefficient in Eq.(\ref{renorm}) does not increase with 
 temperature. From the data of Table~\ref{coeff}, we find that at $T_c$,
 the term $2 b_2^2/c_2$ is still only $10$\% of $a_{41}/4$, less than in
 the 0 K case. These results show that even a large softening of the
 elastic constant $C^\prime$ does not change the character of the phase
 transition, which remains displacive second order.

 Figure~\ref{free-ca} shows also that, above the transition temperature, the
 minimum energy corresponds to the cubic cell and therefore the results
 about the high temperature structural stability of the $c$ phase discussed
 in the previous Section remain valid.

\section{Conclusions}

 The $c \leftrightarrow t$ phase transition of pure stoichiometric
 zirconia has been studied from different theoretical perspectives. Both
 symmetry arguments and the lattice dynamical analysis suggested that
 this transformation might be second-order. In the 0 K perfect $t$
 structure, the primary order parameter $\delta$ has a clear definition
 in terms of the displacement of the oxygen atoms away from their
 centrosymmetric position, which define a zone boundary phonon
 $X_2^-$. But the same definition cannot be directly applied to a finite
 temperature atomic configuration. Instead, we defined the more general
 macroscopic thermodynamic variable as an ensemble average of the
 displacements of the O atoms projected onto the $X_2^-$ zone boundary
 phonon coordinate. The corresponding frequency of vibration was then
 calculated directly from the MD simulations by applying standard
 statistical mechanics techniques.

 The temperature evolution of both the equilibrium order parameter
 $\delta$ and of the corresponding frequency $\nu$ was then followed
 during the MD run.  The results of these simulations have been
 interpreted with the Landau theory: the critical exponent of 0.35 fitted
 to the results is between the extreme values of 0.5 and 0.25
 corresponding to a second-order and to a tricritical phase transition.

 Approaching $T_c$ from the field of stability of the $t$ phase, the
 order parameter $\delta$ gradually decreases up to very near the
 transition point, as in a displacive second-order phase
 transformation. In a temperature window about $T_c$, the large
 fluctuations of the order parameter degrade the quality of the averages
 achievable from MD simulations, but we were able to observe $\nu^2$
 decreasing linearly with $T$ by some 70 \%. In contrast to the
 prediction of the Landau theory, no increase of that frequency in the
 field of stability of the $c$ phase was observed. At high temperatures
 the dynamics of the oxygen sublattice revealed a high degree of mobility
 of the anions and a low correlation between their motion along the
 $X_2^-$ vibrational mode.

 In order to clarify these observations, we calculated the free energy
 surfaces relating the two structures at different temperatures,
 combining constrained MD simulations and the thermodynamic integration
 technique. These calculations showed that the high temperature stability
 of the $c$ structure is due to the entropic contribution $T\Delta S$ and
 not to a variation of the internal energy profile. Moreover, we showed
 that the energy surfaces of the $c$ phase are highly anharmonic, not
 only at the transition temperature, but also well above $T_c$. This
 confirms and may explain in terms of thermodynamic quantities the
 absence of the optical modes of vibration in the experimental spectra of
 Liu {\it et al.},~\cite{Liu87} and the postulated uncorrelated
 ``fluidlike'' motion of the oxygens about their centrosymmetric
 position. Similarly, the experimentally observed increase of the ionic
 conductivity and the ``structural phenomenon'' mentioned by Aldebert and
 Traverse~\cite{Aldebert85} may possibly be connected with the soft
 dynamics of the oxygen sublattice caused by the flat energy surfaces.

 Our analysis revealed the peculiar character of the $c \leftrightarrow
 t$ transformation in zirconia. Approaching $T_c$ from below, the
 variation of the free energy surfaces seems to support the thesis of a
 $t \rightarrow c$ displacive second-order phase transition. On the
 contrary, approaching $T_c$ from above, it is not possible to follow the
 softening of a particular phonon mode. The entropic term $T \Delta S$
 eliminates the double well in the free energy, and frees the atoms to
 move along that mode without an energy cost and without a well defined
 vibrational frequency.  This is more akin to the high temperature phase
 in an order-disorder phase transformation.

\acknowledgements

SF is grateful for support from the European Science Foundation, Forbairt
and the British Council, and for discussions with Nigel Marks. ATP and MWF
are grateful to the EPSRC for funding under Grants No. L66908 and
No. L08380. This work has been supported by the European Communities HCM
Network ``Electronic Structure Calculations of Materials Properties and
Processes for Industry and Basic Science'' under grant No. ERBFMRXCT980178.


%
%

 \begin{table}[b]
 \caption{Order parameters for the $c \leftrightarrow t$ phase
 transition decomposed into irreducible representations of the $O_h$
 cubic group.}  \label{op-irrep} 
 \begin{center}
 \begin{tabular}{l|ll}  
 order parameters & \multicolumn{1}{c}{irrep} \\ \hline 
 $\left( \delta_x, \delta_y, \delta_z \right)$ & & $T_1$ \\ 
 $\,\,\, \eta_1$ & $\left(\epsilon_{xx} + \epsilon_{yy} + \epsilon_{zz} \right)$
 & $A_1$ \\ 
 $(\eta_2,\eta_3)$ & $\left[ \left(2\epsilon_{zz} - \epsilon_{xx} -
 \epsilon_{yy} \right),\sqrt{3}\left(\epsilon_{xx} - \epsilon_{yy} \right) 
\right] $ & $E$ \\ 
$(\eta_4,\eta_5,\eta_6)$ & $ \left( \epsilon_{xy}, \epsilon_{yz},
 \epsilon_{zx} \right)$ & $T_2$ \\  
\end{tabular} 
\end{center}
\end{table}

 \begin{table*}[t]
 \caption{Polynomials in the order parameters of Table~\ref{op-irrep}
 which are invariants under the set of transformations belonging to $O_h$.}
 \label{polnm}
 \begin{center}
 \begin{tabular}{l|l} 
$A_2   (\delta^2)$ & $ (\delta^2_x +\delta^2_y +\delta^2_z )$ \\
$A_{41} (\delta^4)$ & $ (\delta^4_x +\delta^4_y +\delta^4_z )$ \\
$A_{42} (\delta^4)$ & $ (\delta^2_x \delta^2_y 
                       + \delta^2_y \delta^2_z 
                       + \delta^2_z \delta^2_x )$ \\ \hline
$B_{1} (\epsilon, \delta^2) $ & $ (\delta^2_x +\delta^2_y +\delta^2_z )
               (\epsilon_{xx} +\epsilon_{yy} +\epsilon_{zz} )$ \\
$B_{2} (\epsilon, \delta^2) $ & $ (2\delta^2_z - \delta^2_x -\delta^2_y )
                      (2 \epsilon_{zz} - \epsilon_{xx}-\epsilon_{yy} )$ \\ 
   & $ \, + 3 ( \delta^2_x -\delta^2_y ) ( \epsilon_{xx}- \epsilon_{yy} ) $ \\
$B_{3} (\epsilon, \delta^2) $ & $ (\delta_x \delta_y 
                           + \delta_y \delta_z + \delta_z \delta_x )
                (\epsilon_{xy} +\epsilon_{yz} +\epsilon_{zx} )$ \\ \hline 
$C_{1} (\epsilon^2)$ & $ (\epsilon_{xx} +\epsilon_{yy}+\epsilon_{zz} )^2$ \\
$C_{2} (\epsilon^2)$ & $ (2 \epsilon_{zz} - \epsilon_{xx} -\epsilon_{yy}
 )^2 + 3 ( \epsilon_{xx}- \epsilon_{yy} )^2 $ \\
$C_{3} (\epsilon^2)$  & $(\epsilon^2_{xy} +\epsilon^2_{yz}
                                         +\epsilon^2_{zx} )$ \\
 \end{tabular}
 \end{center}
 \end{table*}

 \begin{table}
 \caption{Coefficients for the Landau energy expansion
 (\ref{energy}) at the 0 K equilibrium volume of the $c$ cell: $a_2$ and
 $a_{42}$ in Ry/$a_0^2$, $a_{41}$ in Ry/$a_0^4$, and $c_{2}$ in Ry,
 where $a_0$ is the Bohr radius. The coefficients $b_{1}=-0.363$ Ry/$a_0$
 and $c_{1}=16.768$ Ry complete the 0 K set. See text for the temperature
 dependence of $c_2$.}
 \label{coeff}
 \begin{tabular}{c|ccccc}
 T (K)  & $a_2$   &  $a_{41}$ & $a_{42}$ & $b_{2}$ & $c_{2}$ \\ \hline
0       & -0.0534 &  0.347    & 1.825    & -0.0763  &  1.228 \\ 
50      & -0.0478 &  0.330    & 1.191    & -0.0749  &  1.204 \\
500     & -0.0258 &  0.235    & 0.873    & -0.0705  &  0.998 \\
1000    & -0.0143 &  0.191    & 0.751    & -0.0384  &  0.768 \\
1500    & -0.0058 &  0.184    & 0.661    & -0.0354  &  0.537 \\
2000    &  0.0030 &  0.152    & 0.273    &   ---    &  0.307 
 \end{tabular}
 \end{table}

%
%

\end{document}